\documentclass[showpacs, pra,twocolumn,preprintnumbers ,amsmath, amssymb, superscriptaddress, aps]{revtex4-2}
\usepackage{color}
\usepackage{amsmath,amssymb}
\usepackage{pifont}
\usepackage{amssymb}  
\usepackage{bbold}
\usepackage{makecell}
\usepackage{verbatim}
\usepackage{float}
\usepackage{subfloat}
\usepackage[caption=false]{subfig}
\usepackage{tikz}
\usepackage{makecell}
\usepackage{subfig}
\usepackage{pifont}   
\usepackage{graphicx} 
\graphicspath{{Figures/}}
\usepackage{dcolumn}  
\usepackage{bm}       
\usepackage{multirow} 
\usepackage{placeins}
\usepackage[colorlinks]{hyperref}

\usepackage{mathtools}

\begin{document}
\title{Angle-dependent chiral tunneling in biased twisted bilayer graphene}
\author{Nadia Benlakhouy}
\affiliation{School of Applied and Engineering Physics, Mohammed VI Polytechnic University, Ben Guerir, 43150, Morocco}
\author{El Mustapha Feddi}
\affiliation{School of Applied and Engineering Physics, Mohammed VI Polytechnic University, Ben Guerir, 43150, Morocco}
\author{Abdelouahed El Fatimy}
\affiliation{School of Applied and Engineering Physics, Mohammed VI Polytechnic University, Ben Guerir, 43150, Morocco}
\date{\today}
\begin{abstract}
In twisted bilayer graphene (TBLG), chiral tunneling can be tuned by parameters such as the twist angle, barrier height, and Fermi energy. This differs from the tunneling behavior observed in monolayer and Bernal bilayer graphene, where electrons either pass completely through or are fully blocked due to the Klein paradox. Here we investigate the effect of a perpendicular interlayer bias on electron tunneling through electrostatic barriers in TBLG. Using a dual-gated model, which controls the carrier density and interlayer potential difference independently, we compute the transmission and reflection probabilities of electrons at different angles and energies for representative twist angles of $\theta = 1.8^{\circ}, 3.89^{\circ}, 9.43^{\circ}$. We find that a moderate bias suppresses normal-incidence transmission by opening a band gap in the low-energy spectrum. Our results show this leads to near-total reflection at low energy, with transmission starting to increase just above the gap due to twist-dependent conducting channels. The applied bias breaks the system's effective inversion symmetry, resulting in pronounced direction-dependent and valley-specific asymmetries in the angular distribution of transmitted electrons. We show that electrons incident at different angles show notable variations in transmission under bias. Furthermore, interlayer bias modulates Fabry–Pérot–like resonances in the TBLG barrier, shifting the energies of transmission peaks and altering their intensity.
\end{abstract}
\pacs{ 73.22.Pr, 72.80.Vp, 73.63.-b\\
{\sc Keywords}: twisted bilayer graphene, chiral tunneling, interlayer bias, valley-dependent transport, Fabry–Pérot resonances}
\maketitle
\section{Introduction}
Graphene’s two-dimensional electron gas presents a unique setting for studying the relativistic Klein tunneling paradox, owing to its pseudospin (sublattice) degree of freedom. Charge carriers in a graphene monolayer can transmit perfectly through a potential barrier at normal incidence \cite{katsnelson2006chiral, castro2009electronic, cheianov2006selective, allain2011klein, beenakker2008colloquium}. However, those in a Bernal-stacked bilayer are perfectly reflected under the same conditions \cite{mccann2006landau, katsnelson2006chiral, nilsson2008electronic,  zhang2009direct, cheianov2007focusing}. These two limiting cases, one of perfect Klein tunneling and one of complete pseudospin mediated blocking, are direct consequences of the different chiralities (Berry phases) in monolayer contrary to bilayer graphene \cite{ando1998berry, mccann2013electronic, bliokh2015quantum, young2009quantum, stander2009evidence, xiao2010berry}. Twisted bilayer graphene (TBLG), composed of two misoriented graphene layers, has emerged as a versatile 2D platform in which the twist angle can continuously tune the electronic spectrum and pseudospin configuration \cite{bistritzer2011moire, lopes2007graphene, kim2017tunable}. Notably, at certain magic misalignments, the moiré superlattice of TBLG yields isolated flat electronic bands that host gate-tunable correlated insulators and superconductors, pointing to the rich physics
realized in this system \cite{cao2018unconventional, cao2018correlated, tarnopolsky2019origin, po2018origin, lu2019superconductors, sharpe2019emergent}. Even away from such strongly correlated regimes, TBLG offers remarkable control over Dirac band structure and chirality \cite{bistritzer2011moire, de2011topologically, yoo2019atomic}, enabling precise control over chiral tunneling between monolayer- and bilayer-like regimes
\cite{ kim2024aharonov, amet2012tunneling, liu2020tunable}. A key theoretical study by He {\it et al}. \cite{he2013chiral} demonstrated that TBLG indeed realizes a new incarnation of the Klein paradox. Due to the misaligned layer chiralities, normally incident electrons in an unbiased TBLG do not face a fixed all-or-nothing outcome as in untwisted graphene but instead exhibit an angle and energy tunable probability of transmission \cite{moon2013optical, kim2017tunable, yoo2019atomic}. By adjusting the barrier height or the Fermi energy, one can switch normal incidence tunneling in TBLG from near-perfect transmission to partial or even (near) perfect reflection, and vice versa, thus achieving tunable chiral tunneling that leverages the twist-controlled pseudospin rotation \cite{moon2014electronic, amet2012tunneling, mishchenko2014twist}. Although the analysis of \cite{he2013chiral} was limited to the pristine instance of TBLG with no interlayer potential difference \cite{mccann2006asymmetry, castro2007biased, zhang2009direct, liu2020tunable}, this tunability opens a way for controlling electron flow in 2D systems.

In TBLG, such a bias not only induces a gap in the moiré Dirac bands but also breaks the layer symmetry, leading to new results for an unbiased system \cite{zhang2013valley, jung2011lattice, rickhaus2018transport, sui2015gate}. For example, gapped regions of the moiré pattern can carry opposite valley Chern indices \cite{zhang2013valley, san2013helical, ju2015topological}, and spatially varying bias can give rise to topologically protected zero-line modes \cite{martin2008topological, rickhaus2018transport}. The bias effectively introduces an additional degree of control over the chirality and valley dynamics in TBLG, calling for a systematic study of how it modifies the previously observed chiral tunneling behavior.

Here we investigate chiral tunneling in TBLG under an interlayer bias, focusing on how the angle and energy-dependent transmission and reflection for representative twist angles $\theta=1.8^{\circ}, 3.89^{\circ}, 9.43^{\circ}$ are modified by a perpendicular electric field. Using a dual-gated device configuration in our model, we independently tune the carrier density and interlayer potential difference, analogous to experimental setups in which top and bottom gates control the Fermi level and displacement field \cite{rickhaus2018transport, ju2015topological}. Our calculations reveal that applying a moderate bias has a significant impact on the Klein tunneling characteristics. In the low-energy regime, the bias-induced gap suppresses transmission at normal incidence. However, our results show that transmission is partially increased just above the gap due to twist-induced electronic states that enable tunneling. Moreover, a bias breaks certain symmetries of the TBLG barrier, introducing pronounced angular asymmetries in the tunneling probability \cite{he2013chiral, du2018tuning}. In contrast to the unbiased case, which preserves effective inversion and yields symmetric transmission for $\pm$ incidence angles, a biased TBLG shows direction-dependent tunneling, electrons approaching the barrier at a given angle can transmit with different probabilities depending on their incident direction or valley index, pointing to a bias-tunable chirality and valley filtering effect \cite{li2010observation, rickhaus2018transport}.

Furthermore, interlayer bias has been shown to modulate Fabry-Pérot-like resonances in transmission spectra, shifting the energies and narrowing or widening the resonant peaks, providing an external means of tuning quantum interference patterns in electron transport \cite{varlet2014fabry}. Our results clarify our understanding of pseudospin- and valley-dependent tunneling in moiré systems and have significant implications for device engineering. Recent experiments have achieved fine control of carrier density and displacement field in TBLG devices, making the above effects directly testable \cite{ju2015topological, rickhaus2018transport}. The bias controlled suppression or enhancement of transmission at select angles and energies could be exploited in directional tunneling transistors, where the current can be switched on/off by reorienting the incidence or toggling the bias, in principle enabling large on/off ratios that overcome graphene’s usual limitation of zero band gap \cite{sui2015gate}. At the same time, the bias-induced inversion-symmetry breaking and resulting valley asymmetry suggest the possibility of valley-selective electron beam splitters or filters, for example, an electrostatic barrier in TBLG can selectively direct $K$-valley vs. $K'$-valley electrons into opposite layers or angles \cite{jung2011lattice, li2010observation}, and engineered three-way junctions in minimally twisted bilayers have been proposed as perfect valley filters controllable by gating, with potential for creating electron beam splitters and interferometers \cite{martin2008topological}. By quantifying the bias dependence of chiral tunneling in TBLG, our work provides key metrics, such as tunable transmission anisotropy and energy-resolved tunneling contrast, that highlight the potential of TBLG  for next generation electronic and valleytronic devices.

The paper is organized as follows. In Sec. \ref{S1}, we present the continuum model and transfer matrix formalism used to describe chiral tunneling in biased TBLG. In Sec. \ref{S2}, we analyze the angular, energy, and barrier-height dependence of transmission and reflection for different twist angles, focusing on the role of interlayer bias in suppressing Klein tunneling and inducing valley-selective asymmetries. Finally, in Sec. \ref{S3}, we summarize our main results and discuss their implications for directional tunneling devices and valleytronic applications.
\section{Model and method}\label{S1}
\begin{figure}[htp]
    \centering
    \subfloat[]{\includegraphics[width=0.45\linewidth]{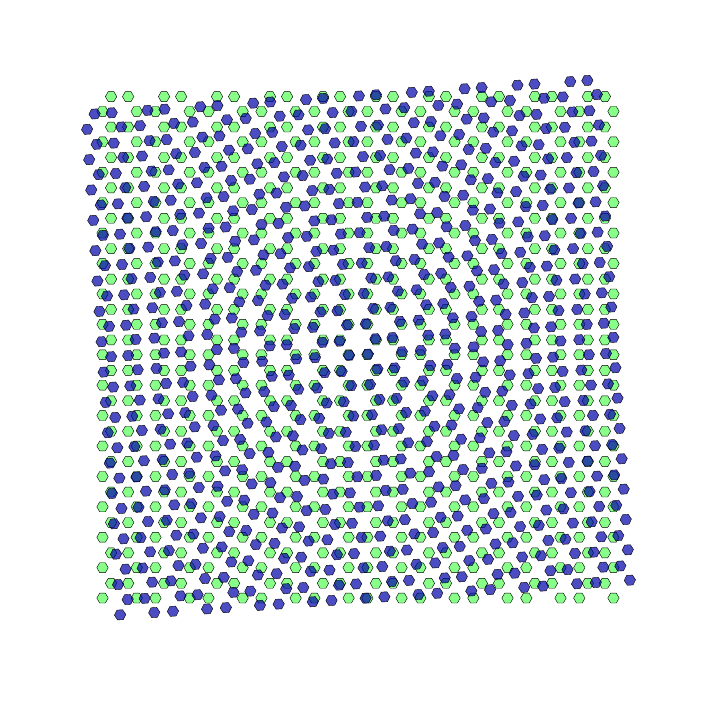}}\hfill
    \subfloat[]{\includegraphics[width=0.45\linewidth]{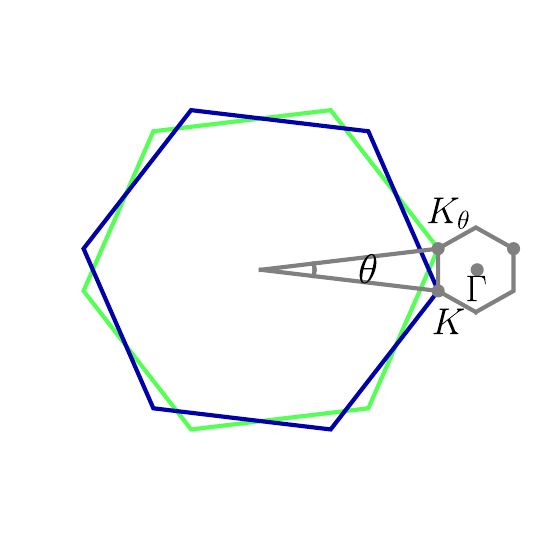}}\\[1.2ex]
    
    \subfloat[]{\includegraphics[width=0.48\linewidth]{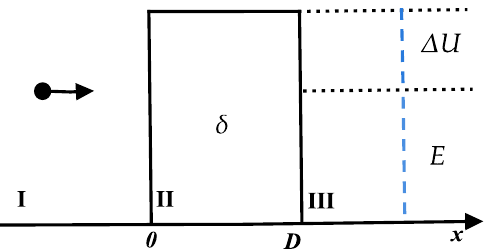}}\hfill
    \subfloat[]{\includegraphics[width=0.28\linewidth]{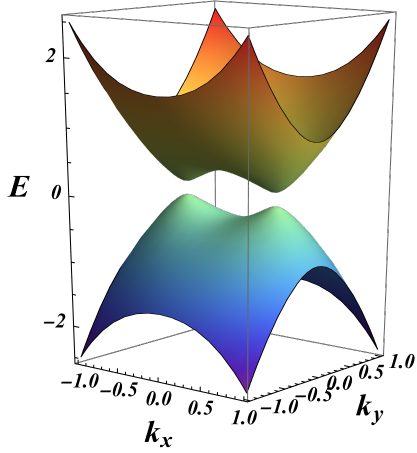}}
    
    \caption{(Color online). Structural and electronic properties of twisted bilayer graphene (TBLG). (a) Schematic of TBLG formation by relative rotation $\theta = 3.89^{\circ}$ between the top and bottom graphene layers. (b) Hexagonal moiré Brillouin zone (MBZ) resulting from periodic superlattice modulation. (c) Schematic of electron tunneling through a one-dimensional potential barrier of height $E + \Delta U$ and width $D$, where $E$ is the Fermi energy. The barrier extends infinitely along the $y$-direction. (d) Bias-dependent energy spectrum ($\delta = 0.1$ eV) showing a bandgap opening at the Dirac points due to symmetry breaking.}
    \label{TBLG system}
\end{figure}
To describe the low-energy electronic spectrum of TBLG, we employ a continuum model based on the nearest-neighbor tight-binding approximation, as introduced in \cite{lopes2007graphene, lopes2012continuum, de2011topologically, choi2011angle}
\begin{equation}\label{Heff}
\mathcal{H}^{\text{eff}} = \gamma
\begin{pmatrix}
\delta/\gamma & \left(k_\theta^{\dagger}\right)^2 - \left(k^{\dagger}\right)^2 \\
k_\theta^2 - k^2&-\delta/\gamma
\end{pmatrix}
\end{equation}
where  $\gamma=\left(2 v_F^2/15 \tilde{t}_{\perp}\right)^2$ is the effective interlayer coupling parameter. Here  $\tilde{t}_{\perp} \simeq 0.4 t_{\perp}$ accounts for the twist-induced reduction in interlayer coupling \cite{lopes2007graphene, lopes2012continuum}, 
$\delta$ corresponding to an externally induced interlayer potential difference.
$k=k_x+i k_y$ and $k^{\dagger}=$ $k_x-i k_y$ are the in-plane wave vector and its conjugate, respectively, with $k_{x, y}=-i \partial_{x, y}$. For a twist angle $\theta \neq 0$, we can define the complex wave vector in the MBZ and its conjugate as
\begin{align}
&k_\theta=\Delta K / 2=\left(\Delta K_x+i \Delta K_y\right) / 2,\\
& k_\theta^{\dagger}=\Delta K^{\dagger} / 2=\left(\Delta K_x-i \Delta K_y\right) / 2
\end{align}
For simplicity, we set $\Delta K_x=0$ and $\Delta K_y=\Delta K$, where $\Delta K=4\pi\sin{(\theta)}/3a_0$, and $a_0=0.25$ nm. This continuum description is valid when the lattice constant is much smaller than the moiré superlattice period $L$, a condition that holds for small twist angles $\theta\leq 10^{\circ}$. Van Hove singularities (VHS) have been observed experimentally in TBLG for twist angles up to $\theta= 10^{\circ}$, a regime where the continuum model is used as a perfect low-energy approximation \cite{he2013chiral, ohta2012evidence, li2010observation, luican2011single, yan2012angle, brihuega2012unraveling, meng2012enhanced}. These singularities arise from saddle points in the band structure formed by the overlap of Dirac cones from the two rotated layers. In the continuum model, the energy of these VHSs can be approximated by
\begin{equation}
E_{\text{V}} = \pm\frac{\hbar v_F |\Delta \mathbf{K}| - 2t_\perp}{2}
\end{equation}
where $\Delta \mathbf{K}$ is the momentum shift between the Dirac points and $t_\perp$ is the interlayer coupling. For the twist angles considered in this work, the VHS appears at $E_V \approx 0.03$ eV for $\theta = 1.8^\circ$, $E_V \approx 0.15$ eV for $\theta = 3.89^\circ$, and $E_V \approx 0.88$ eV for $\theta = 9.43^\circ$, in agreement with scanning tunnelling spectroscopy and theoretical predictions \cite{ li2010observation, luican2011single, brihuega2012unraveling, huang2017evolution}. These values establish the characteristic energy scale for moiré miniband formation and determine the beginning of strong interlayer hybridization. Notably, the proximity of the VHS to the Fermi level is significant in identifying transmission and reflection spectra in our study. At low twist angles, where it $E_V$ is small, the Fermi energy lies near the VHS even for modest doping, enhancing the density of states and enabling strong quantum interference. This results in sharp Fabry–Pérot-like resonances and increased sensitivity to interlayer bias. In contrast, for large angles $\theta = 9.43^\circ$, the high value of $E_V$ places the VHS far from the Fermi level under moderate gating, and the tunneling behavior closely resembles that of decoupled monolayers, with weak interlayer hybridization and broad non-resonant transmission \cite{li2010observation, luican2011single, ohta2012evidence}.

However, an interlayer bias $\delta$ breaks this symmetry, shifts the VHS in energy, and modifies the density of states by enhancing the curvature near the band edges. This significantly alters chiral tunneling, instead of perfect transmission at normal incidence, the bias induces a gap and pseudospin mismatch, leading to strong reflection and angle-dependent tunneling suppression. These effects are consistent with observations in both biased TBLG and other systems where external fields or strain control VHS behavior \cite{li2010observation, bi2019designing}.
By solving the eigenvalue problem of Eq.~\eqref{Heff}, we obtain the energy dispersion relation that captures the interplay between the twist-induced moiré pattern, interlayer coupling, and the external bias
\begin{widetext}
		\begin{equation}\label{energyeq}
		\epsilon\left(k_x, k_y\right)= \pm \frac{1}{\gamma} \sqrt{\delta^2+\gamma^2\left[\left(k_x^2-k_y^2-\frac{1}{4} \Delta K_x^2+\frac{1}{4} \Delta K_y^2\right)^2+\left(2 k_x k_y-\frac{1}{2} \Delta K_x \Delta K_y\right)^2\right]}
	\end{equation}
\end{widetext}
where $\epsilon=E-U$ is the energy relative to the potential 
$U(x)$. To study the scattering of Dirac fermions in TGB through a simple barrier structure along the $x$-direction, we apply a rectangular potential barrier in region II
 \begin{equation}
U(x) = 
\begin{cases}
0, & x < 0 \quad \text{(Region I)} \\
E + \Delta U, \quad & 0 < x < D \quad \text{(Region II)} \\
0, & x > D \quad \text{(Region III)}
\end{cases}
\end{equation}
$\Delta U$ is the energy difference between the height of the potential barrier $U$ and the incident energy $E$ (see Fig. \ref{TBLG system} (c)).
Due to the translation invariance in the $y$-direction, the solution to Eq. \eqref{Heff} in $j$-th region can be written as $\psi_j(x, y)= \psi_j(x) e^{i k_y y}$. The wave functions $\psi_j(x)$ in the three regions $x<0$,
 $0<x<D$ and $x>D$ are written as follows
	\begin{widetext}
	\begin{equation}
	\begin{aligned}
		& \psi_I(x)=\frac{1}{\sqrt{2}}\binom{1}{s_1 \rho_1^{+}} e^{i k_{x 1} x}+\frac{b_1}{\sqrt{2}}\binom{1}{s_1 \rho_1^{-}}e^{-i k_{x 1}x}
		 +\frac{c_1}{\sqrt{2}}\binom{1}{s_1 \rho_2^{-}} e^{k_{x2} x}, \\
		& \psi_{II}(x)=\frac{a_2}{\sqrt{2}}\binom{1}{s_2 \zeta_1^{+}} e^{i q_{x 2} x}+\frac{b_2}{\sqrt{2}}\binom{1}{s_2 \zeta_1^{-}} e^{-i q_{x 2} x}
		+\frac{c_2}{\sqrt{2}}\binom{1}{s_2 \zeta_2^{+}} e^{q_{x2} x}+\frac{d_2}{\sqrt{2}}\binom{1}{s_2 \zeta_2^{-}} e^{-q_{x 2} x}, \\
		& \psi_{III}(x)=\frac{a_3}{\sqrt{2}}\binom{1}{s_3 \rho_1^{+}} e^{i k_{x_{1}} x}+\frac{c_3}{\sqrt{2}}\binom{1}{s_3 \rho_2^{+}} e^{-k_{x_{2}} x},
	\end{aligned}
		\end{equation}
	\end{widetext}
	where	
	\begin{equation}
\rho_j^{ \pm}= \frac{\left( \pm k_{x j}+i k_y\right)^2-
k_\theta^2}{\left|\left( \pm k_{x j}+i k_y\right)^2-k_\theta^2\right|},\quad \zeta_j^{ \pm}=\frac{\left( \pm i q_{x j}+i k_y\right)^2-k_\theta^2}{\left|\left( \pm i q_{x j}+i k_y\right)^2-k_\theta^2\right|.}
	\end{equation}
with
$s_{1,3}=\operatorname{sgn}\left(\epsilon\right) ,s_2=\operatorname{sgn}\left(\epsilon-U\right),$ 
$ k_{x1}$ is the positive real root and $ik_{x2}$ is the positive imaginary root of Eq. (\ref{energyeq}) in the $I$, $III$ regions.
 The coefficients $a_j, b_j, c_j,$ and $d_j$, $j=1,2,3$ are determined by matching both components of the spinor and their derivatives at the interfaces $x=0$ and $x=D$. A full transfer‐matrix derivation and the explicit form of $\mathbf{M}$ are given in Appendix \ref{App.A}.
 
\section{Results and Discussion}\label{S2}
\begin{figure}[ht]
    \centering
    \subfloat[$\theta=1.8^{\circ}$]{\includegraphics[width=0.45\linewidth]{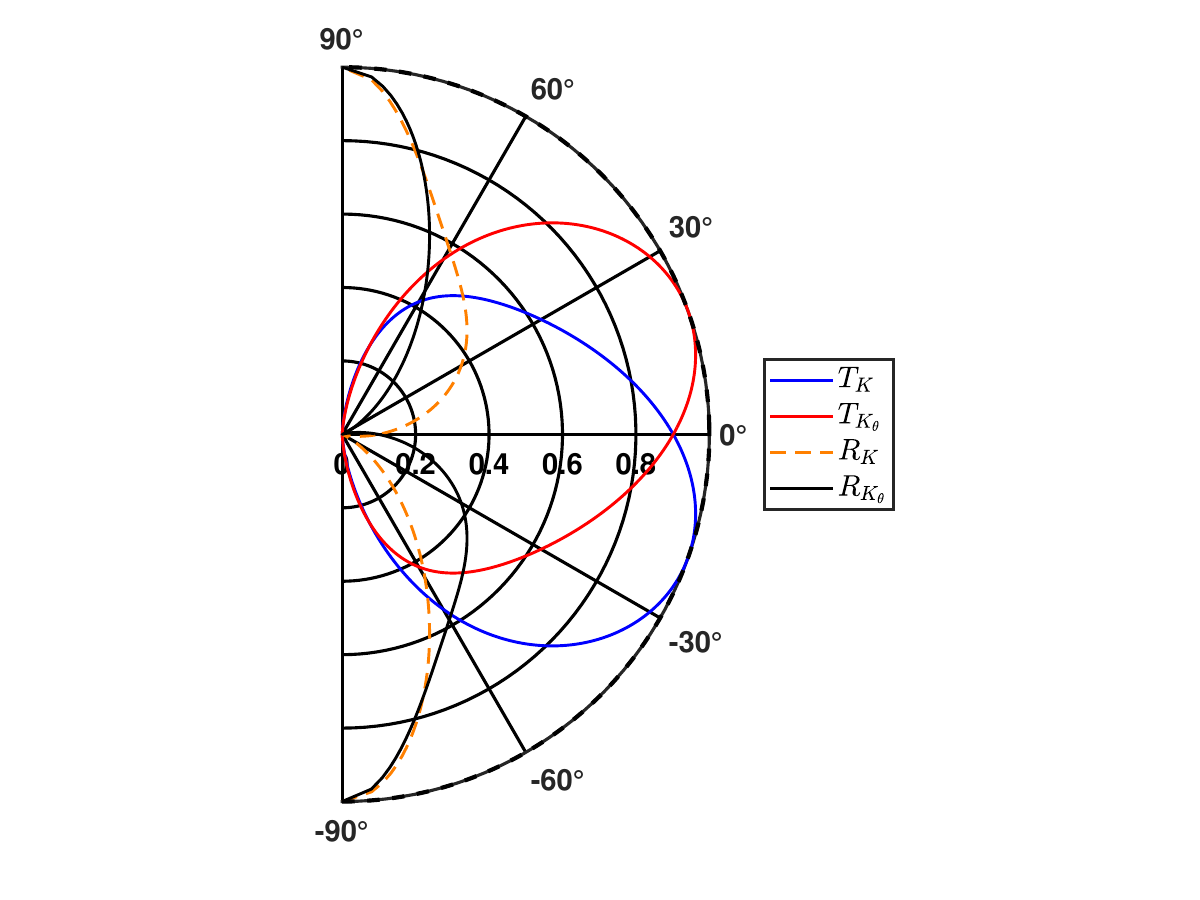}}\hfill
    \subfloat[$\theta=3.89^{\circ}$]{\includegraphics[width=0.45\linewidth]{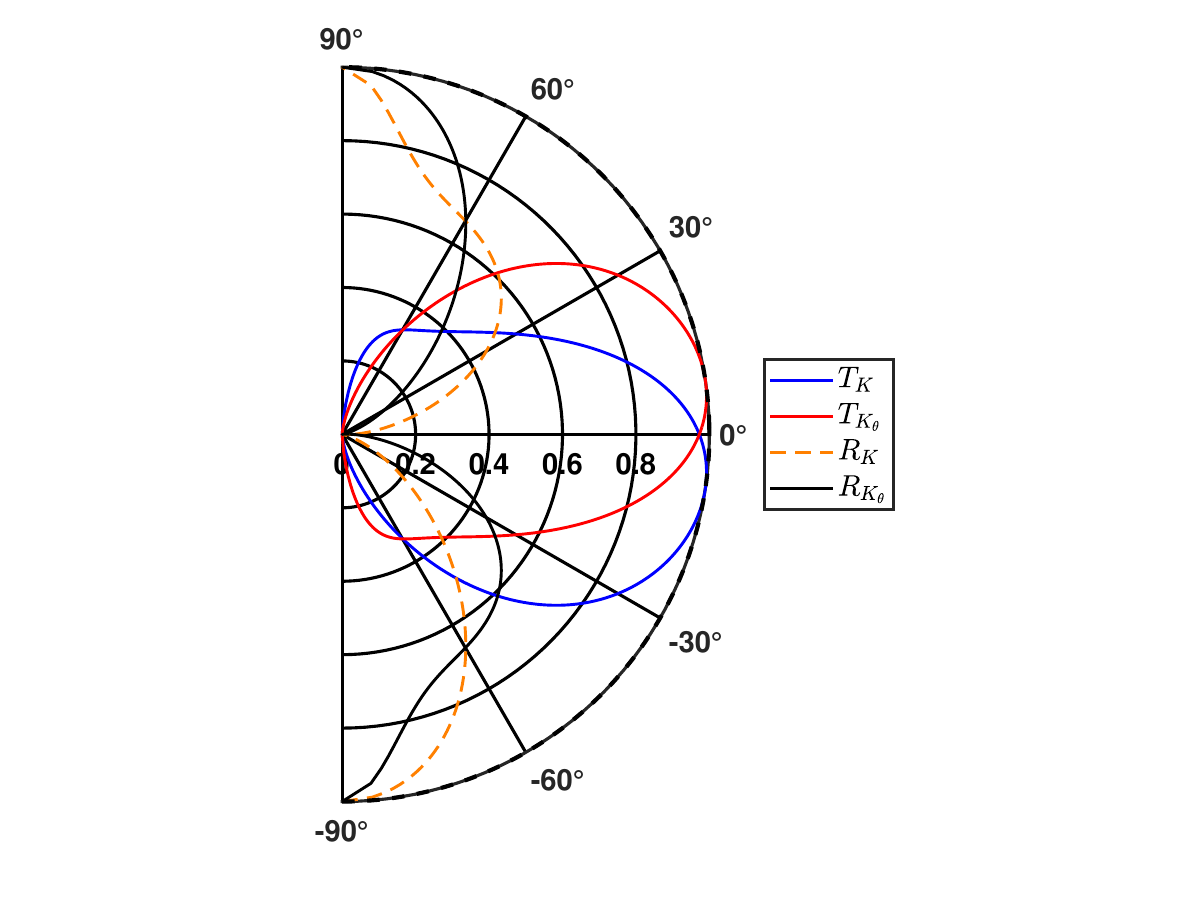}}\\[1ex]
    
    \subfloat[$\theta=9.43^{\circ}$]{\includegraphics[width=0.45\linewidth]{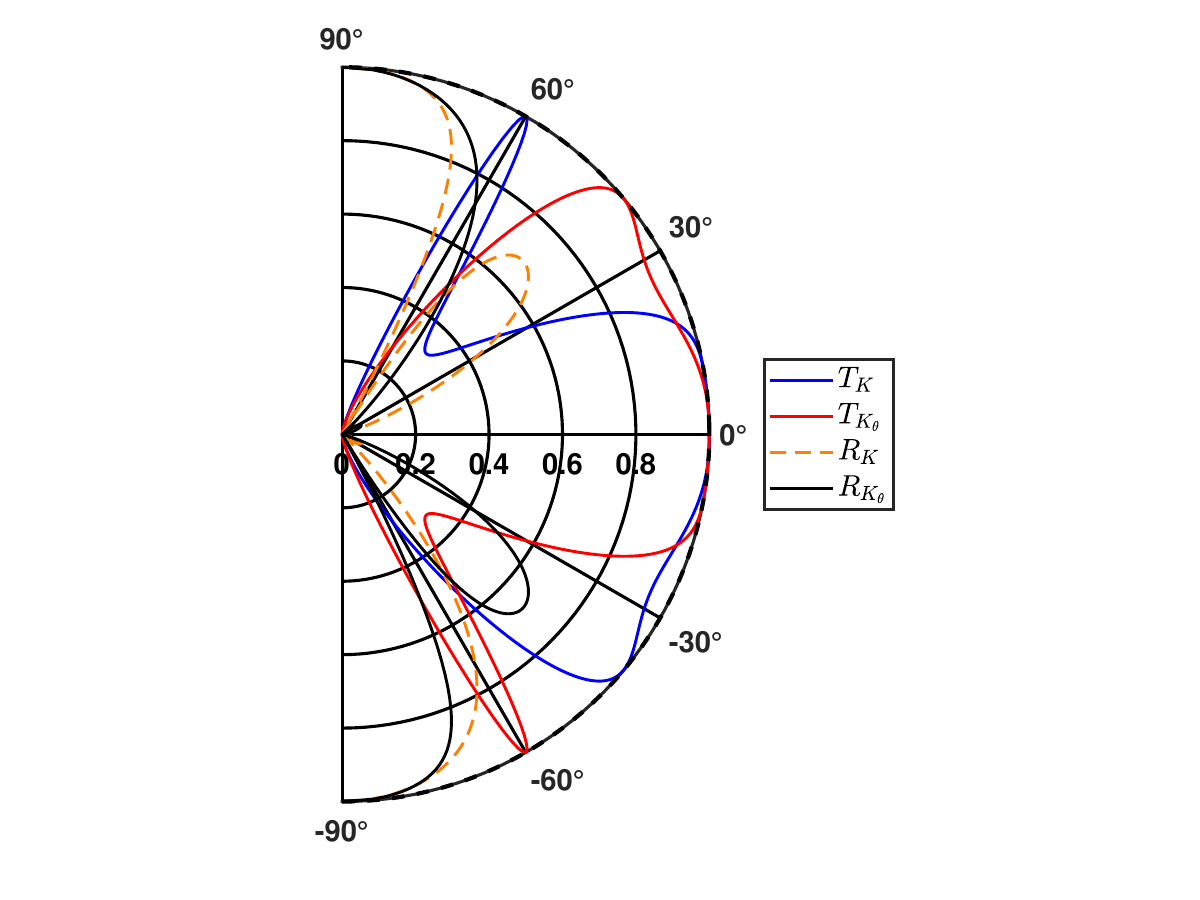}}\hfill
    \subfloat[$\theta=1.8^{\circ}$]{\includegraphics[width=0.45\linewidth]{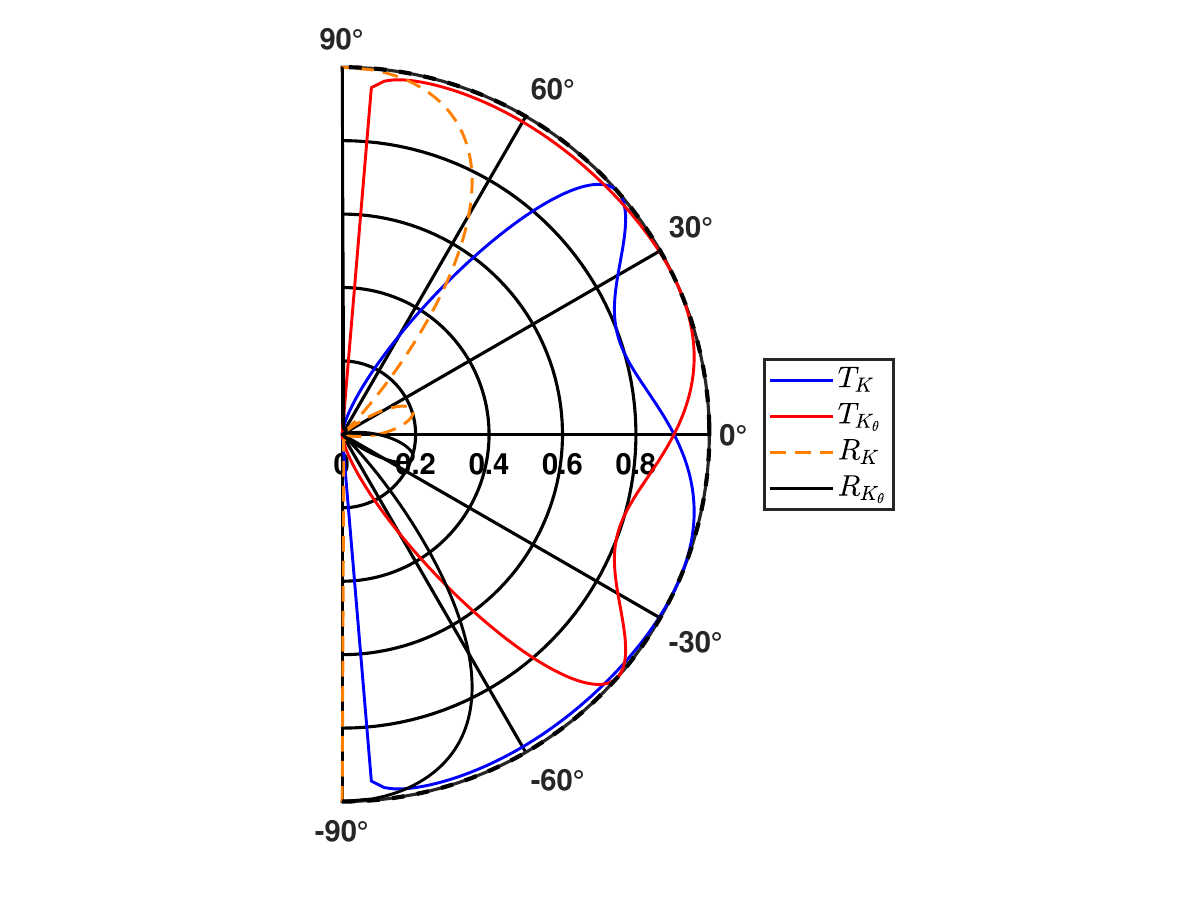}}\\[1ex]
    
    \subfloat[$\theta=3.89^{\circ}$]{\includegraphics[width=0.45\linewidth]{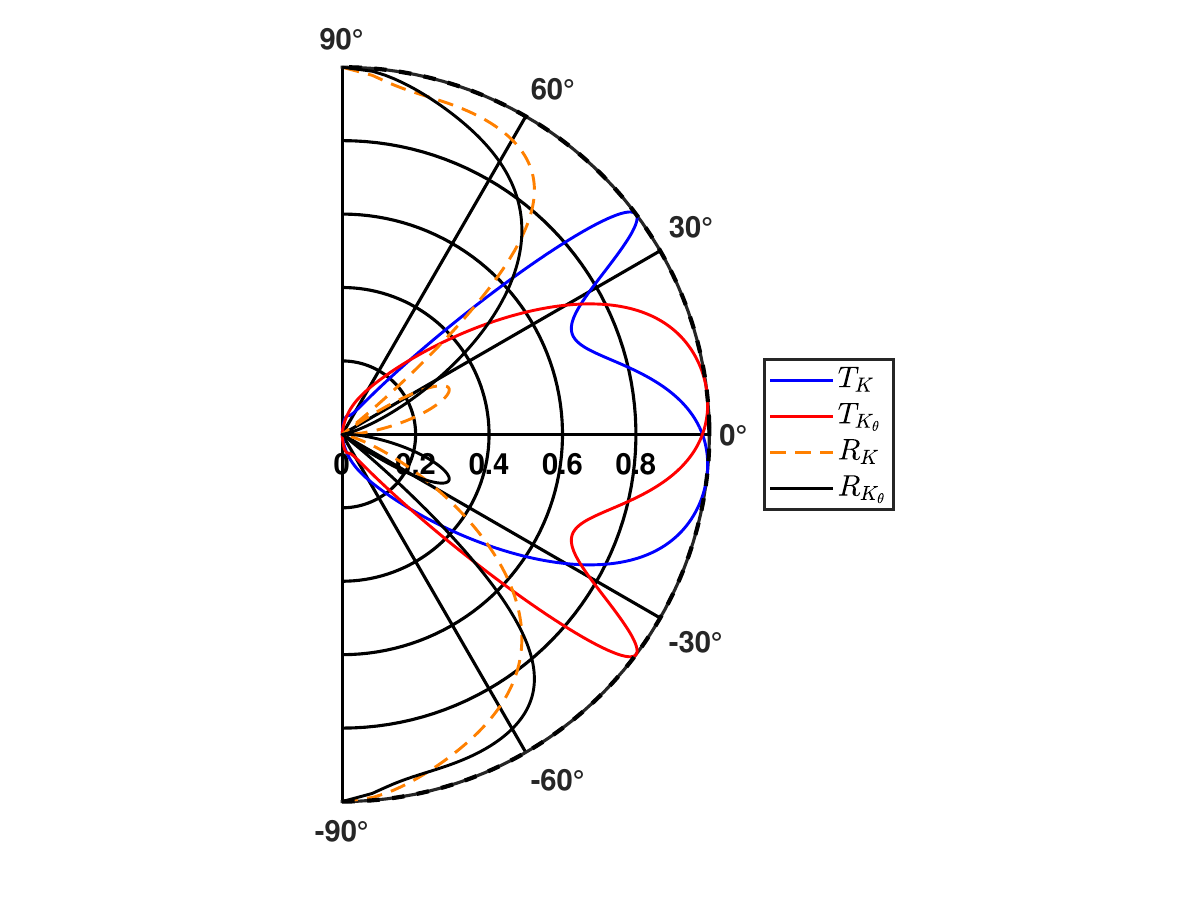}}\hfill
    \subfloat[$\theta=9.43^{\circ}$]{\includegraphics[width=0.45\linewidth]{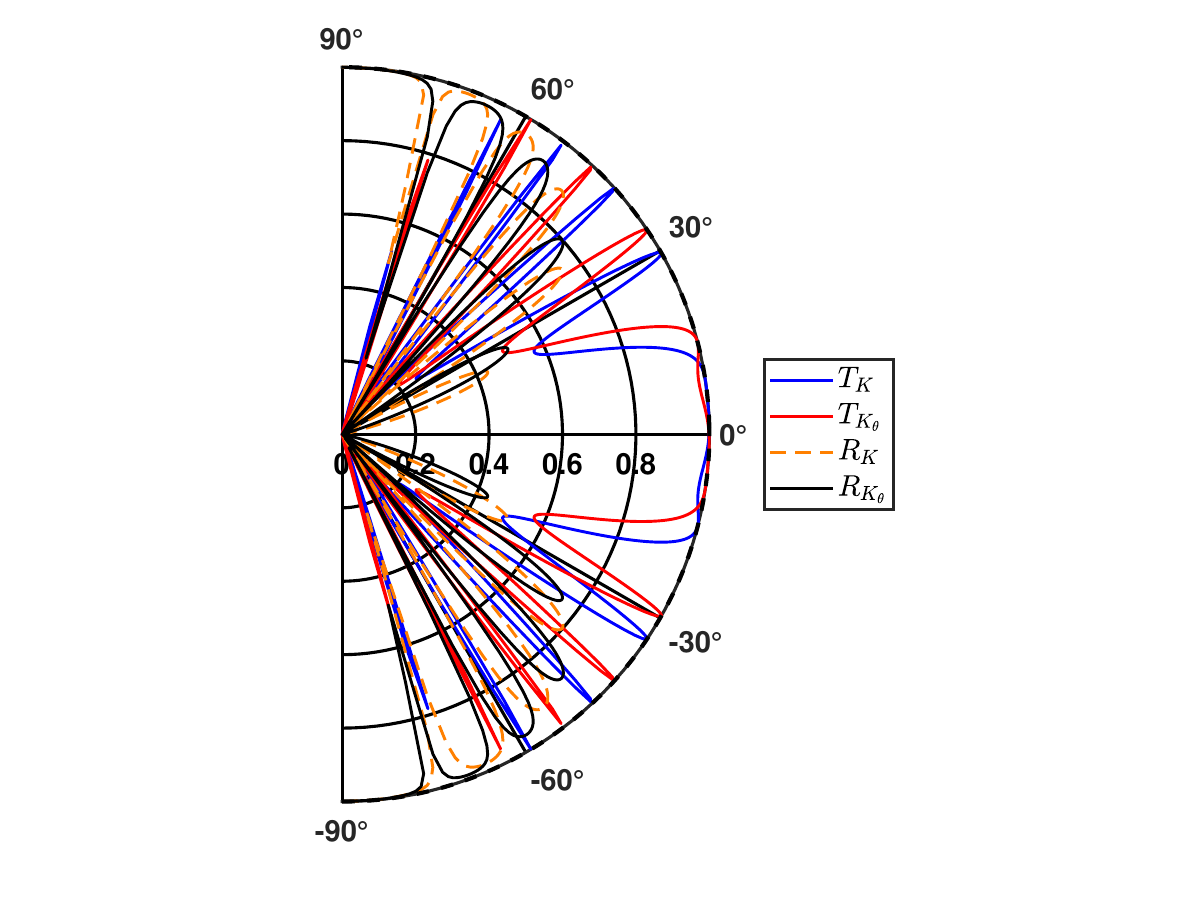}}
    \caption{ (color online). Angle-dependent transmission and reflection probabilities through a 100 nm wide potential barrier as functions of the incident angle $\phi$, for both $K$ and $K_\theta$ cones in biased twisted bilayer graphene. The remaining parameters are $\Delta U = 0.05$ eV, $\delta = 0.2$ eV, and $E/E_V = 0.2$ for panels (a), (b), and (c), and $E/E_V = 0.4$ for panels (d), (e), and (f).}
    \label{T1}
\end{figure}

In Fig. \ref{T1}, we plot the angle-resolved transmission, $T(\phi)$, and reflection, $R(\phi)$, probabilities for biased TBLG at twist angles $\theta = 1.8^\circ$, $3.89^\circ$, and $9.43^\circ$, under a fixed interlayer bias of $\delta = 0.2$ eV and barrier height $\Delta U = 0.05$ eV, these three angles demonstrate a progressive shift from bilayer to monolayer-like tunneling behavior as predicted by the continuum model \cite{lopes2007graphene}. In contrast, biasing shows valley and direction selectivity. At $\theta = 1.8^\circ$ (Figs. \ref{T1} (a-d)), strong moiré coupling and the bias-induced gap suppress normal-incidence transmission to values below 0.1, leading to nearly total reflection $R(0)\gtrsim 0.9$. Off-normal transmission is enhanced, with sharp lobes appearing at 
$|\phi|\approx 60^\circ$, both $K$ and $K_\theta$
valleys exhibit nearly mirror-symmetric angular profiles in 
$T(\phi)$ and $R(\phi)$, resembling the behavior of biased Bernal bilayer graphene, where Klein tunneling is strongly suppressed \cite{katsnelson2006chiral}. 
At $\theta = 3.89^\circ$ (Figs. \ref{T1} (b-e)) a reduced pseudospin mismatch enhances the transmission $T(0)\approx 0.3$ and lowers $R(0)\approx 0.6$, shifting transmission peaks to $\phi\approx\pm45^\circ$ with complementary reflection minima, the broken inversion symmetry now causes opposite lobe enhancement and reflection dips in the two valleys, allowing valley-polarized by moiré-induced Dirac-point asymmetries. \cite{ohta2012evidence}. At $\theta = 9.43^\circ$ (Figs. \ref{T1} (c-f)), weak interlayer coupling makes the bias gap negligible, $T(0)$ rises to $\simeq0.6$, $R(0)$ falls to $\simeq0.4$, and broad transmission lobes center at $|\phi|\lesssim 30^\circ$, closely approximating monolayer Klein tunneling yet exhibiting moderate anisotropy akin to single-layer breakdown effects \cite{luican2011single}. With increasing twist angle 
$\theta$, the pseudogap narrows across all incident angles, while transmission peaks shift toward normal incidence $\phi=0$, leading to enhanced $T(0)$ and reduced 
$R(0)$. This evolution maps a transition from angle-selective filtering at small $\theta$ to nearly isotropic transmission at larger twists.
 In comparison to the \cite{he2013chiral} study, our biased TBLG shows that breaking the layer symmetry not only modulates the magnitude of normal-incidence tunneling but also lifts the mirror symmetry between valleys, leading to angle-dependent transmission peaks that are valley-specific. These observations imply that a transverse electric bias can turn a twisted bilayer into a valleytronic lens or filter, where, for example, electrons from one valley are preferentially transmitted at $+\phi$. At the same time, those from the other are favored at $-\phi$.
\begin{figure}[ht]
    \centering
    \subfloat[$\theta=1.8^{\circ}$]{\includegraphics[width=0.45\linewidth]{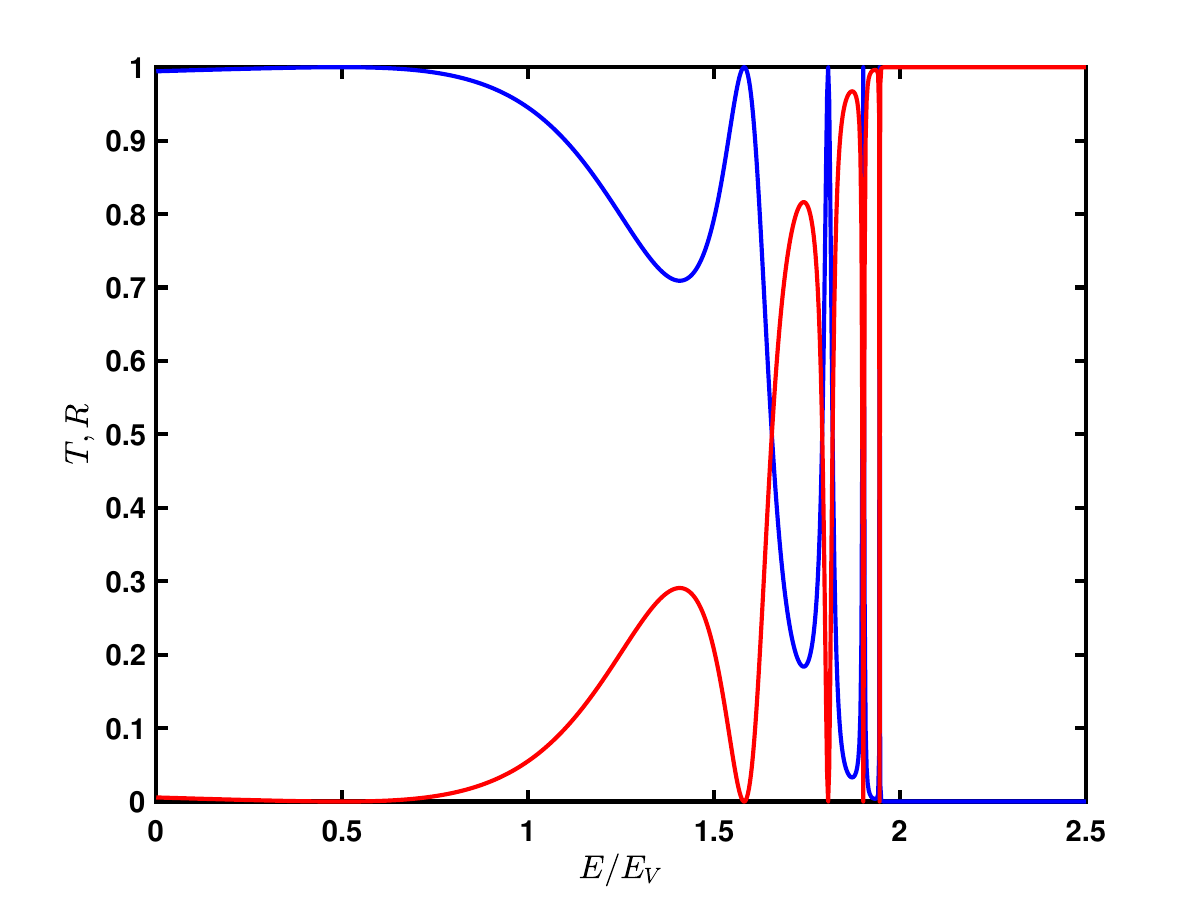}}\hfill
    \subfloat[$\theta=3.89^{\circ}$]{\includegraphics[width=0.45\linewidth]{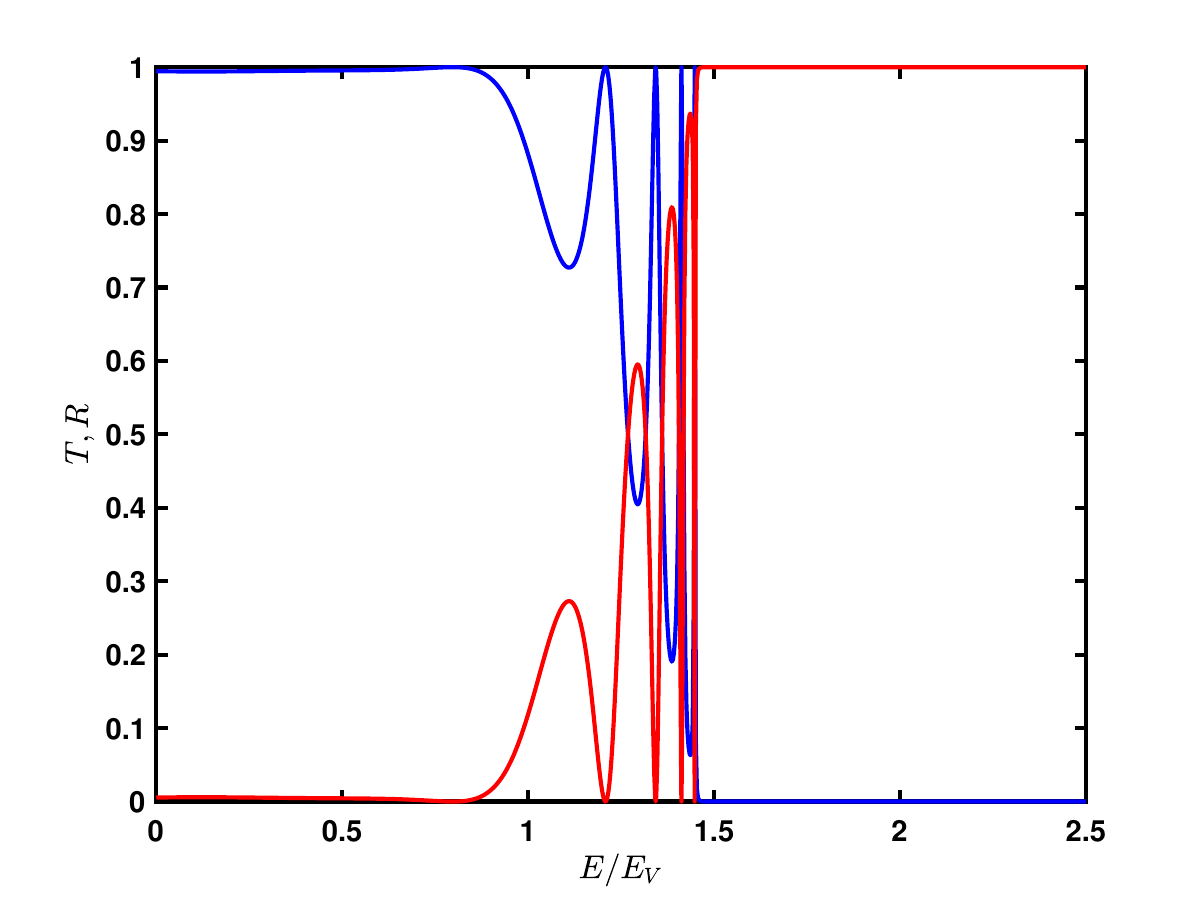}}\\[1ex]
    
    \subfloat[$\theta=9.43^{\circ}$]{\includegraphics[width=0.45\linewidth]{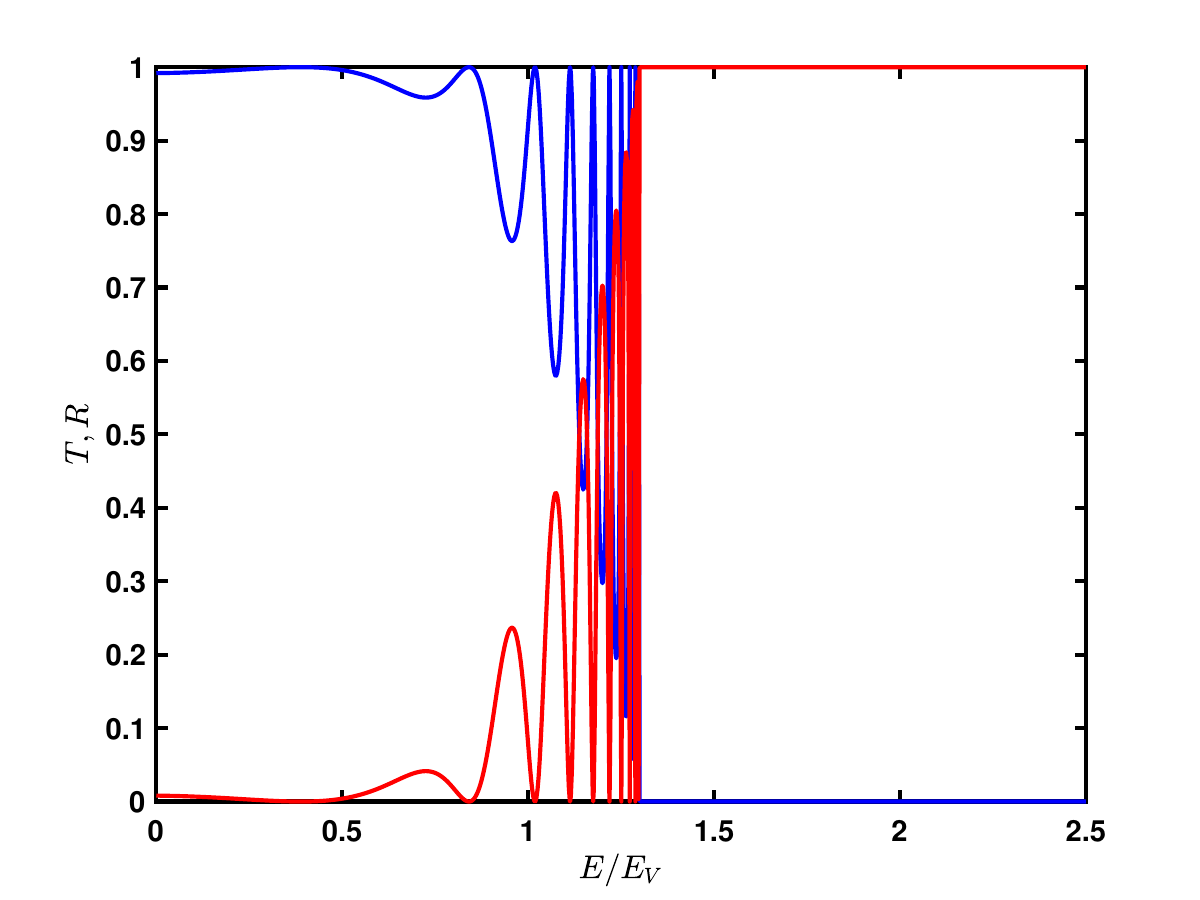}}\hfill
    \subfloat[$\theta=1.8^{\circ}$]{\includegraphics[width=0.45\linewidth]{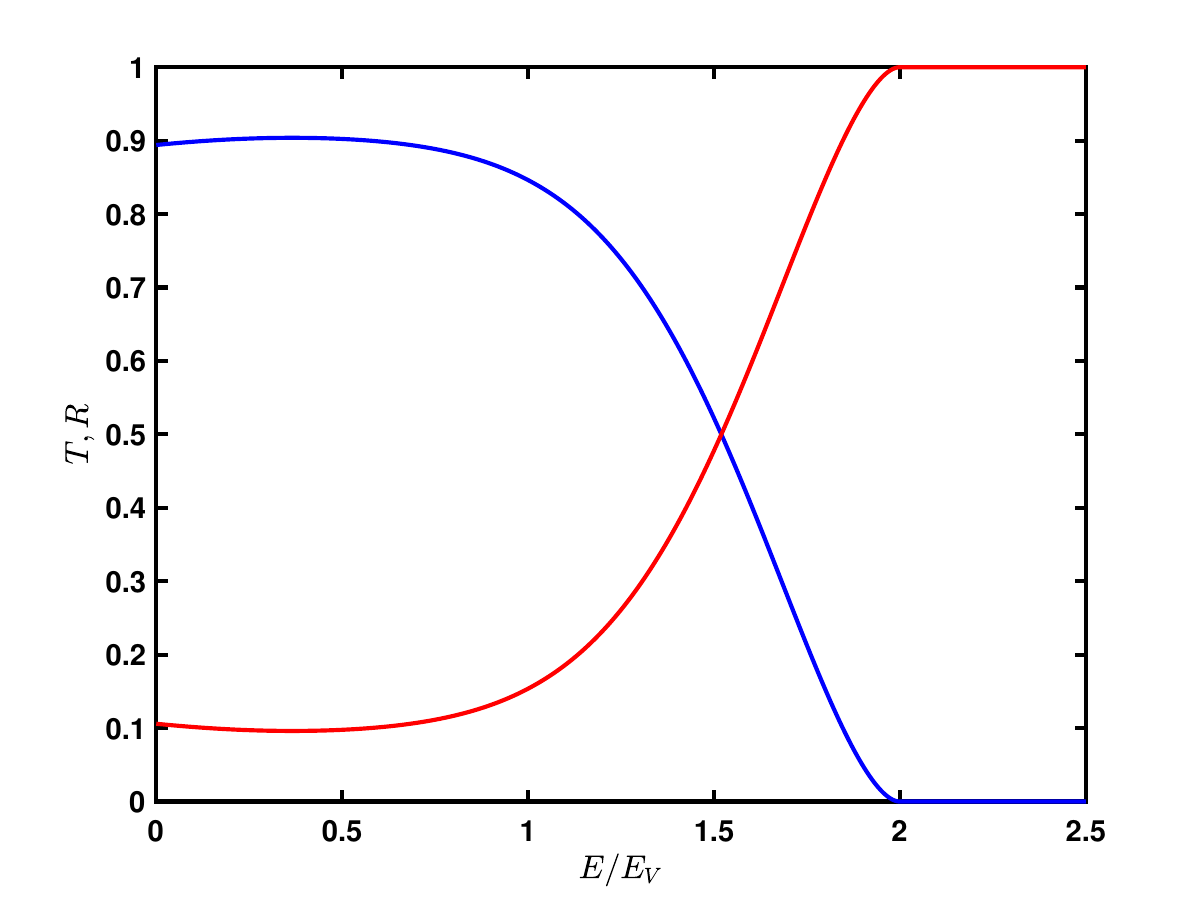}}\\[1ex]
    
    \subfloat[$\theta=3.89^{\circ}$]{\includegraphics[width=0.45\linewidth]{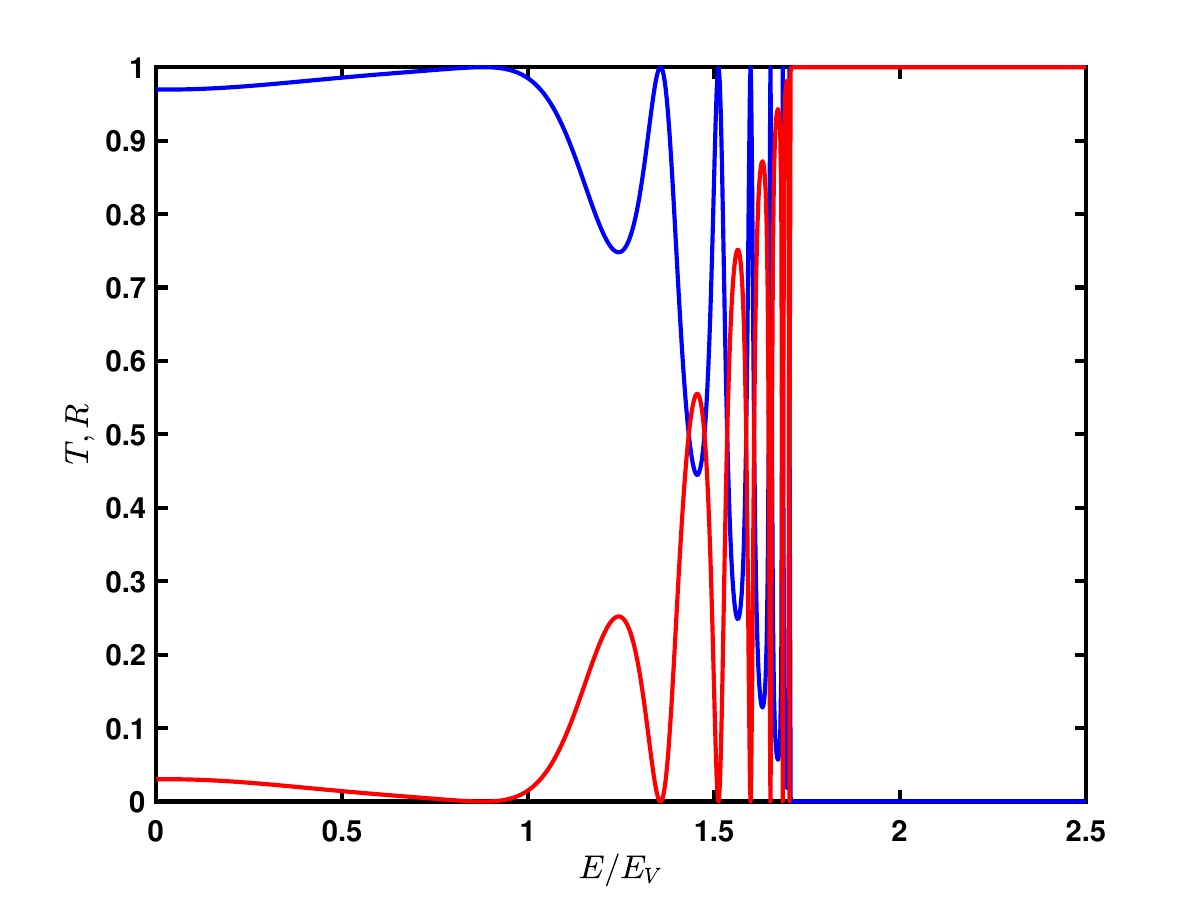}}\hfill
    \subfloat[$\theta=9.43^{\circ}$]{\includegraphics[width=0.45\linewidth]{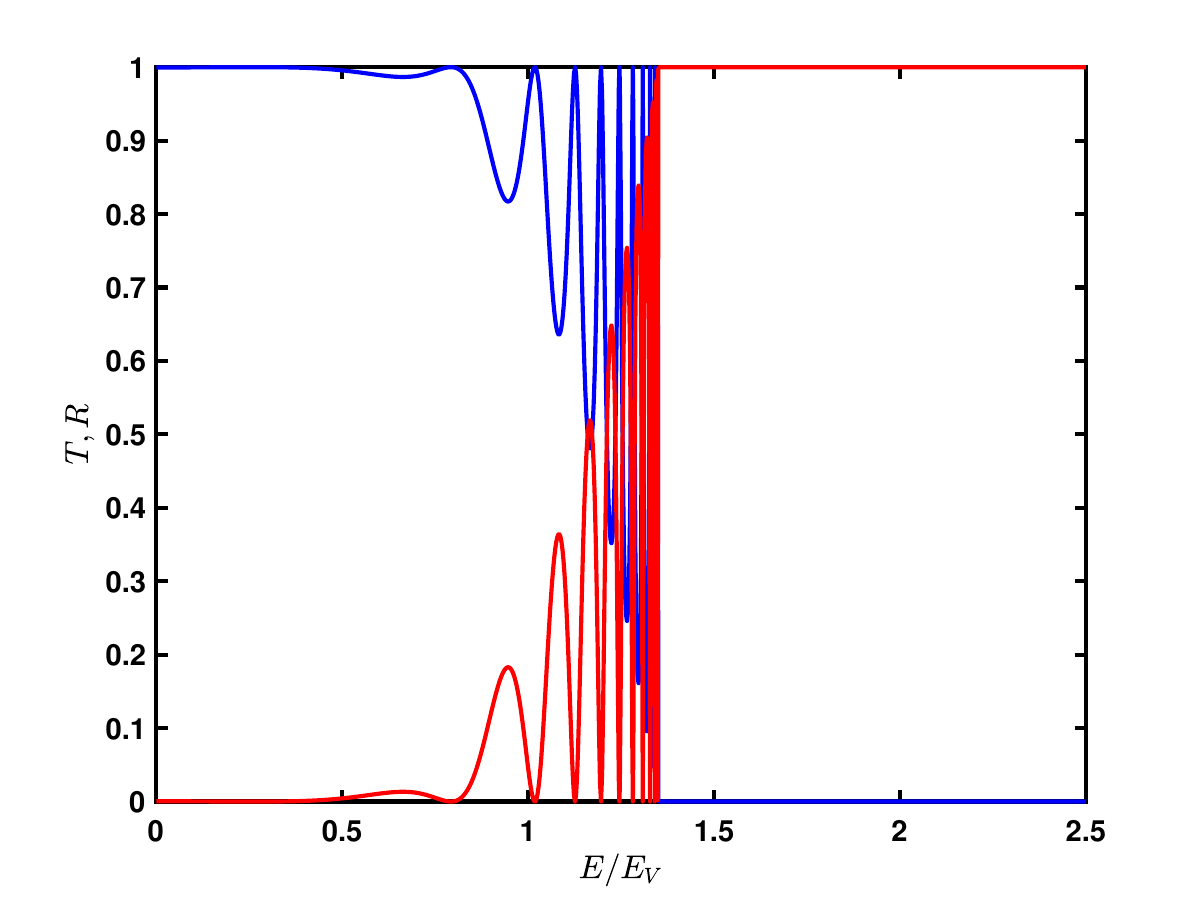}}
    \caption{ (color online). Transmission and reflection probabilities through a 100 nm wide potential barrier as functions of the incident energy $E/E_V$ in biased twisted bilayer graphene. The remaining parameters are $\delta = 0.2$, eV, and barrier height $\Delta U = 0.02$ eV for panels (a), (b), and (c), and  $\Delta U = 0.05$ eV for panels (d), (e), and (f).}\label{T2}
\end{figure}
		\begin{figure}[ht]
    \centering
    \subfloat[$\theta=1.8^{\circ}$]{\includegraphics[width=0.45\linewidth]{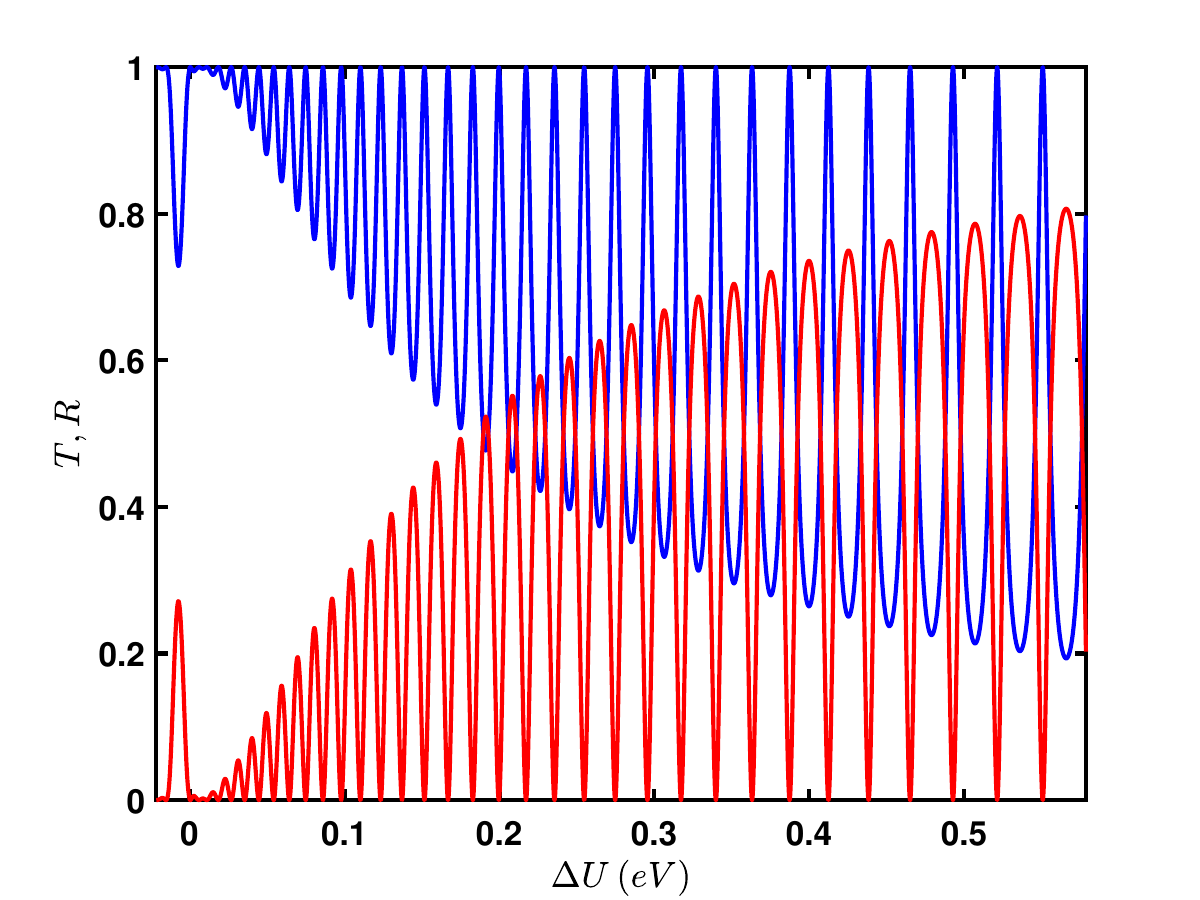}}\hfill
    \subfloat[$\theta=3.89^{\circ}$]{\includegraphics[width=0.45\linewidth]{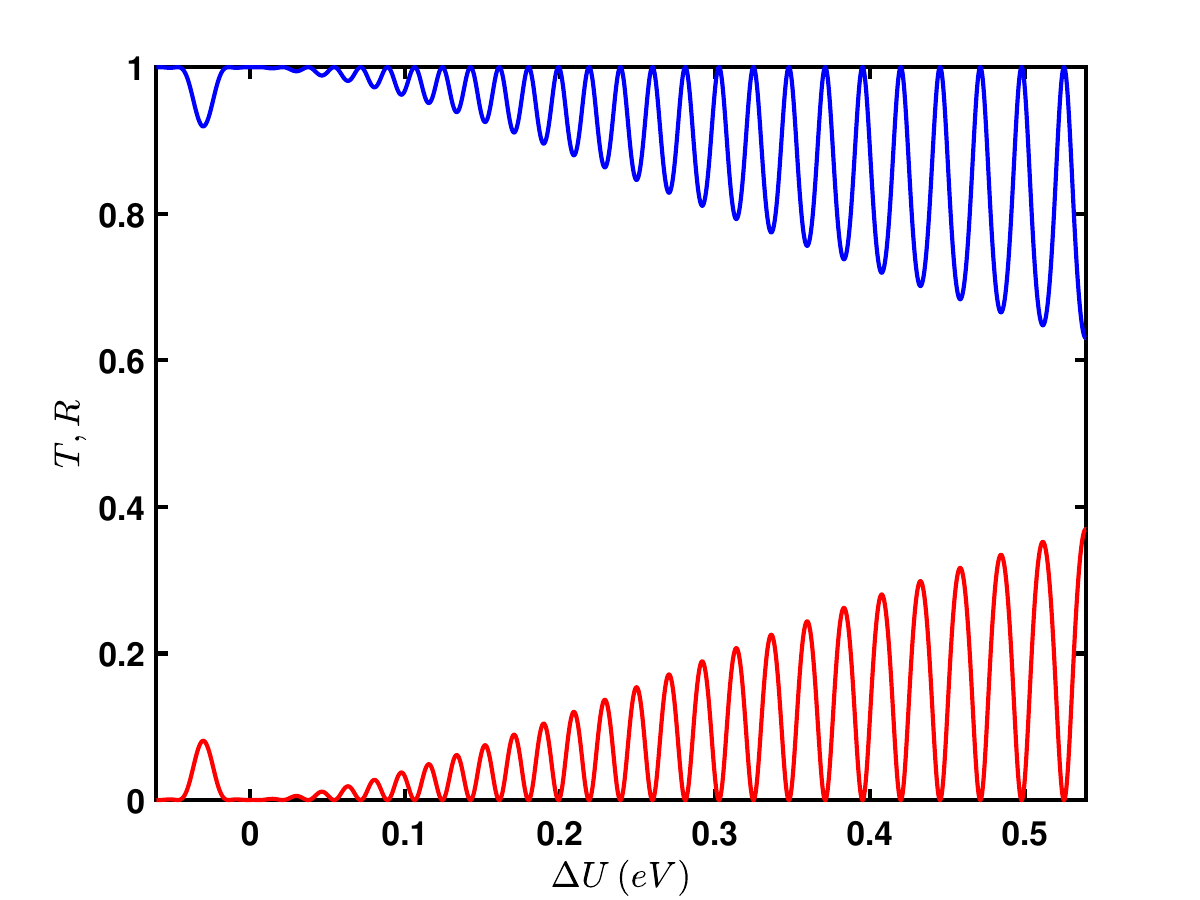}}\\[1ex]
    
    \subfloat[$\theta=9.43^{\circ}$]{\includegraphics[width=0.45\linewidth]{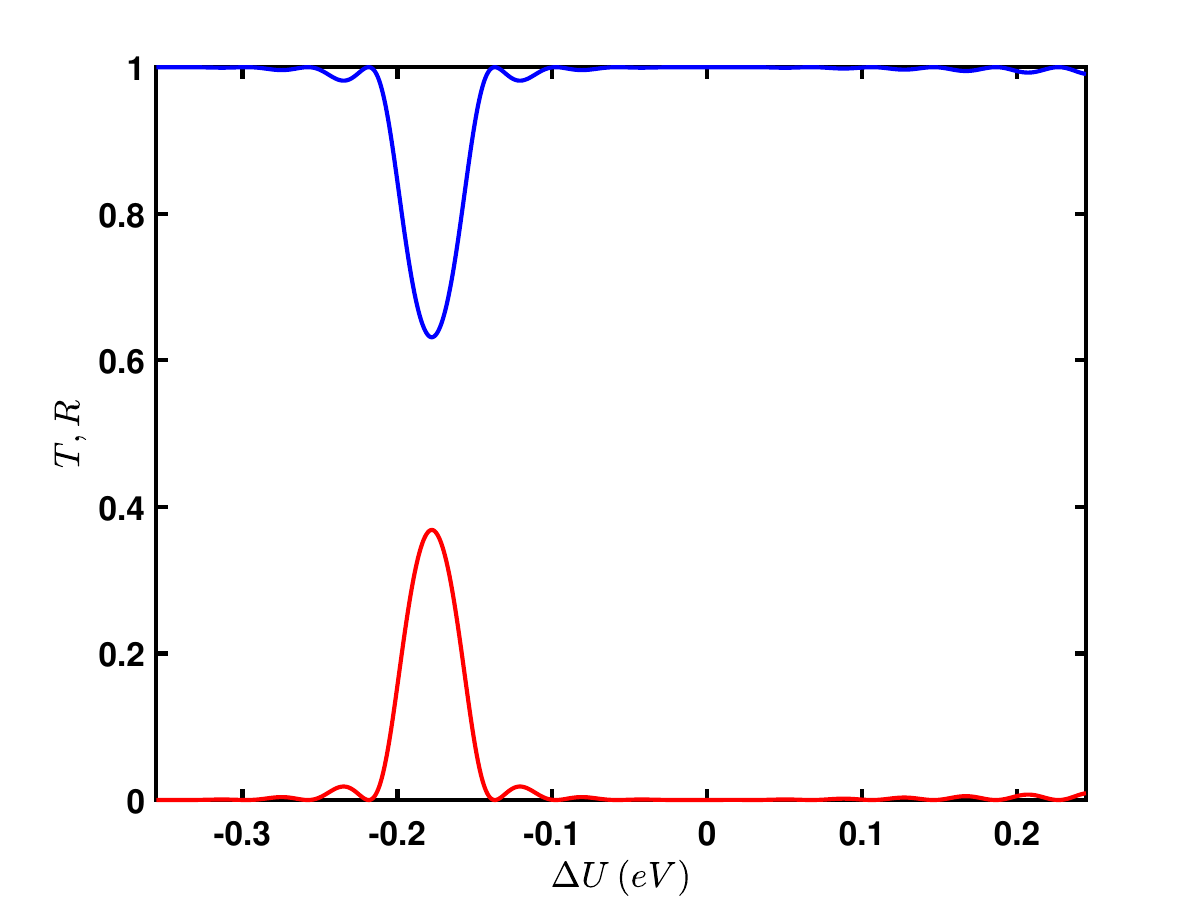}}\hfill
    \subfloat[$\theta=1.8^{\circ}$]{\includegraphics[width=0.45\linewidth]{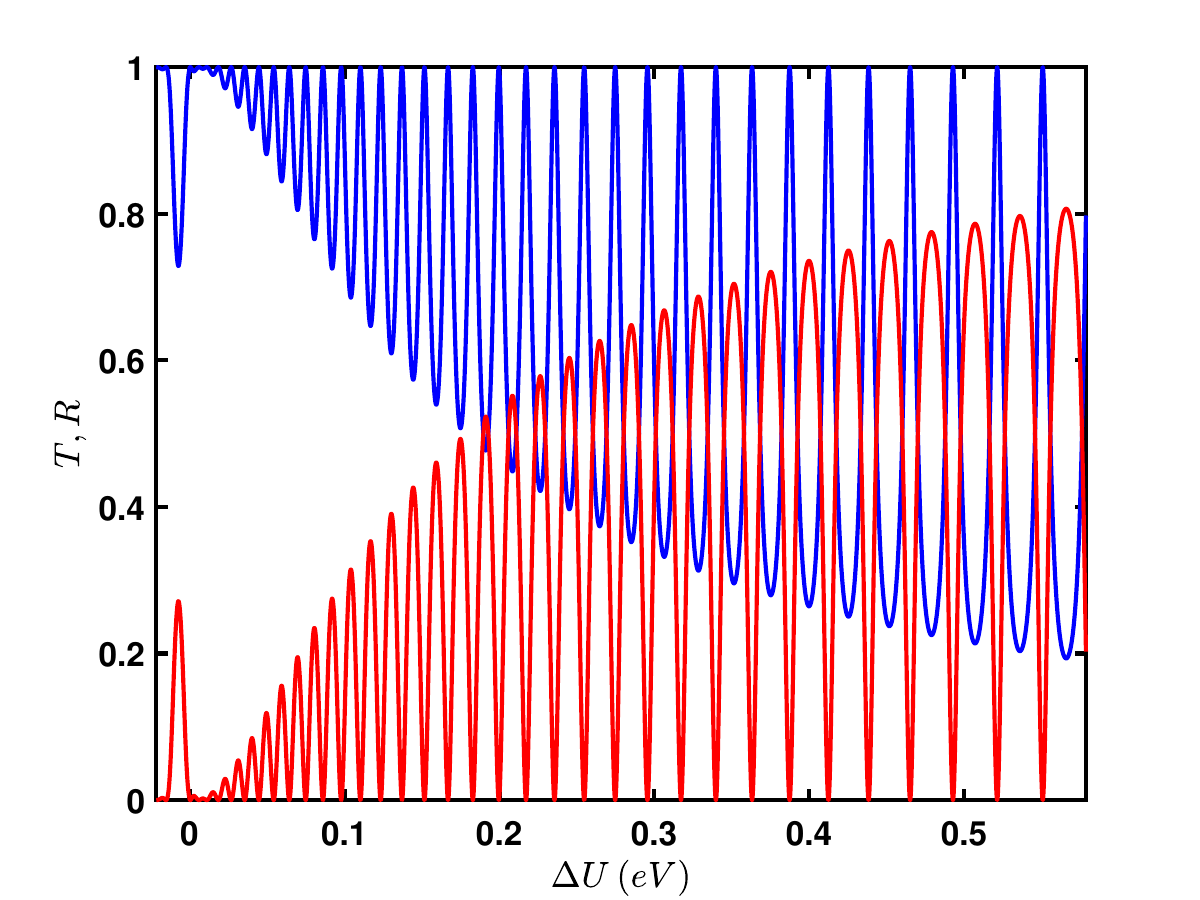}}\\[1ex]
    
    \subfloat[$\theta=3.89^{\circ}$]{\includegraphics[width=0.45\linewidth]{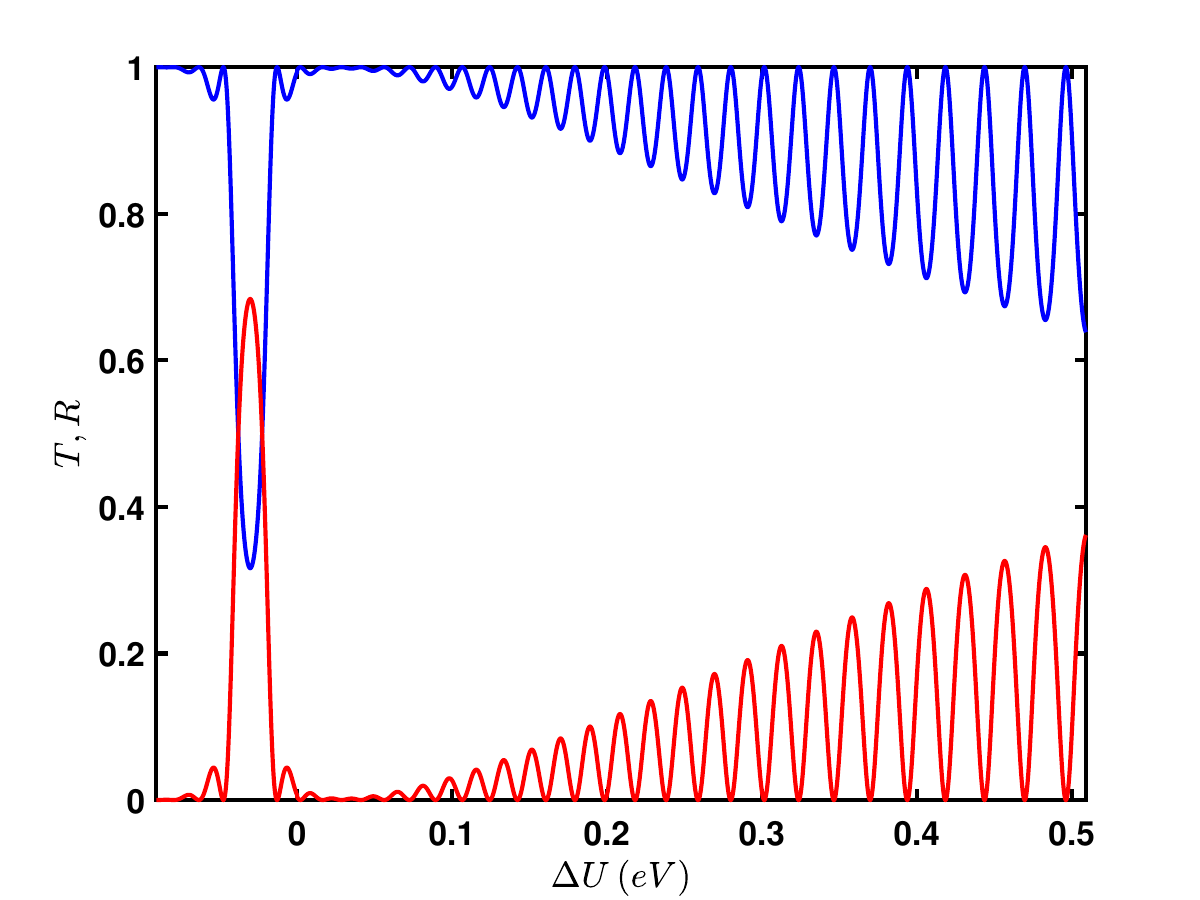}}\hfill
    \subfloat[$\theta=9.43^{\circ}$]{\includegraphics[width=0.45\linewidth]{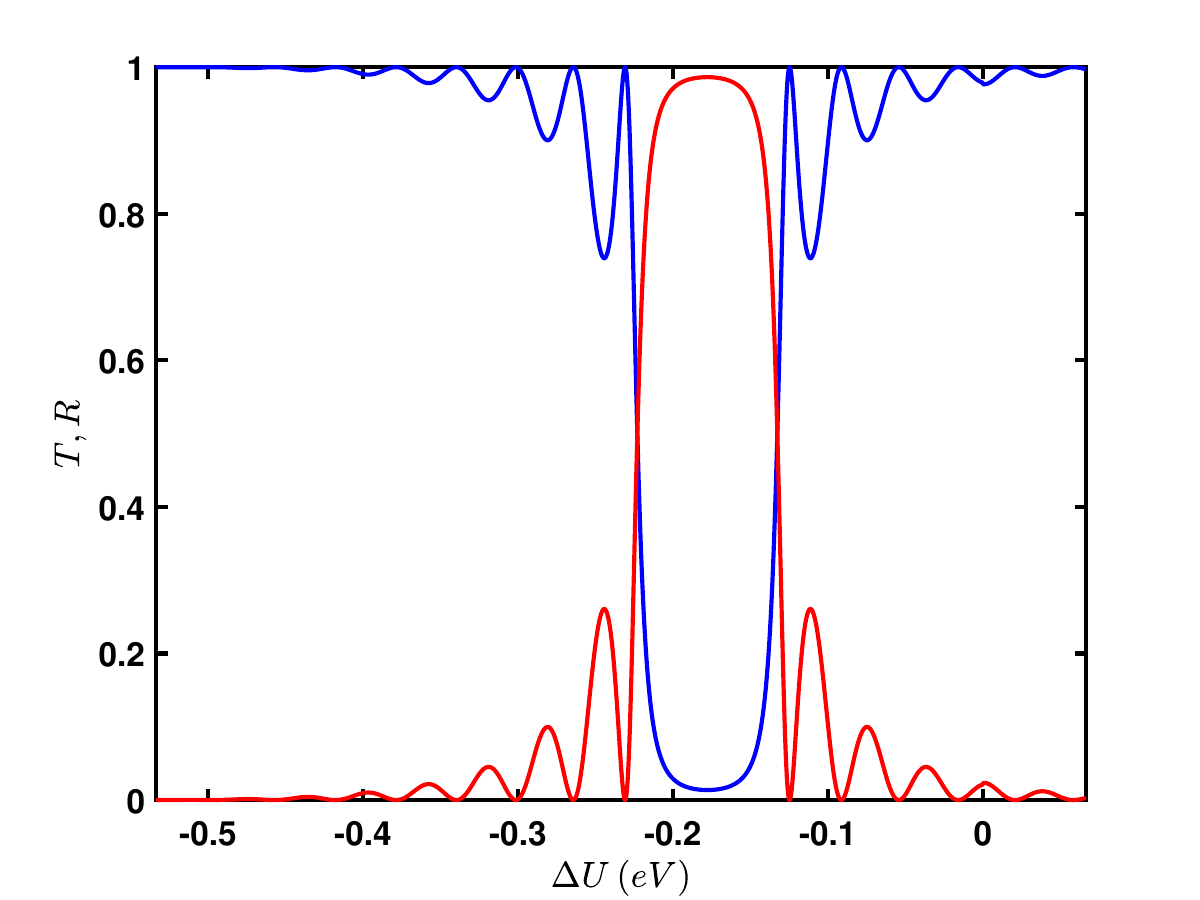}}
\caption{ (color online). Transmission and reflection probabilities through a $100$ nm wide potential barrier as functions of the barrier height $\Delta U$, in biased twisted bilayer graphene. The remaining parameters are interlayer  $\delta = 0.2$ eV, with $E/E_V = 0.2 $ eV for panels (a), (b), and (c), and $E/E_V= 0.4$ eV for panels (d), (e), and (f).}\label{T3}
\end{figure}
 As shown in Figs. \ref{T2} (d–f), the energy-dependent transmission in biased TBLG shows pronounced resonances that evolve with $E/E_V$ and vary strongly with twist angle. Notably, at the smallest twist $\theta = 1.8^\circ$, the transmission starts near unity as $E/E_V=0$, an unexpected result given that a perpendicular bias typically opens a significant band gap in bilayer graphene up to $\sim 0.2$ eV that would ordinarily suppress low-energy tunneling \cite{zhu2017edge}. This near-perfect low-energy transmission under bias can be attributed to the moiré superlattice enabling alternate conducting channels specifically, at small $\theta$ bias, the AB/BA stacked regions into gapped states with opposite valley Chern numbers, but topologically protected helical modes occurs along the domain boundaries \cite{san2013helical}, preserving a high conductance reminiscent of Klein tunneling. As $E/E_V$ increases, both $T$ and $R$ develop a series of resonances associated to twist-dependent band structure, around $E/E_V\approx1$ the van Hove singularity energy where the two layers’ Dirac bands intersect \cite{huang2017evolution}, $T(E)$ exhibits noticeable dips with corresponding $R$ peaks, indicating enhanced backscattering due to the high density of states and reduced group velocity near these saddle points. At higher energies approaching $\sim2E_V$, the system enters the regime where, in the absence of bias, one expects a pseudospin-$1$, this behavior effectively suppresses tunneling at normal incidence, \cite{he2013chiral} showed that pristine TBLG transitions from perfect transmission at low energy to near-total reflection beyond $\sim2E_V$. In the presence of an interlayer bias, the transmission becomes less effective, and increases at higher energies. This behavior reflects the shift and splitting of resonance conditions induced by the bias, which reduces the strength of destructive interference. As a result, $T$ does not drop fully to zero and occurs at modified energies compared to the unbiased case. Comparing across twist angles, the bias induced modulation of resonances is most pronounced at large twist angles, for $\theta=9.43^{\circ}$ smallest moiré period, the layer potential difference effectively decouples the layers at $E\ll E_V$, and only once the incident energy overcomes the $\sim 0.2$ eV interlayer overlap do significant transmission peaks reappear, at $\theta=3.89^{\circ}$ an intermediate behavior is seen, with partial low-energy transmission and moderately strong resonances. Overall, interlayer bias functions as a key control parameter to tune the tunneling spectrum of TBLG, combining monolayer-like Klein transparency with gapped bilayer-like reflection properties. Even with a fixed barrier $\Delta U=0.05$ eV, that alone would produce Fabry–Pérot–type oscillations in graphene, the addition of bias and the twist induced van Hove singularities combine to yield a rich structure in $T(E)$ and $R(E)$ from near-unity transmission at low energies for small twist angles to sharp resonance dips at higher energies, $E/E_V$ in agreement with the findings for pristine TBLG but now precisely shifted and reduced by the applied bias.

In Fig. \ref{T3} (a–c) we plot the normal incidence transmission, $T(\Delta U)$, and reflection, $R(\Delta U)$, probabilities through a $100$ nm barrier as functions of barrier height $\Delta U$ at $E/E_V=0.2$ eV for twist angles $\theta=1.8^\circ$, $3.89^\circ$, and $9.43^\circ$. At $\theta=1.8^\circ$, $T$ starts at unity for $\Delta U=0$ and then exhibits strong Fabry–Pérot–type oscillations of gradually decreasing amplitude as $\Delta U$ increases, reflecting multiple quasibound state resonances inside the bias‐induced gap, in line with gapped bilayer resonances in Bernal graphene \cite{bai2007klein}. For $\theta=3.89^\circ$, the moiré coupling is weaker, giving smaller amplitude oscillations and a more gradual decay of $T$ with $\Delta U$, reminiscent of the biased, energy‐controlled resonant tunneling described by \cite{he2013chiral}. By contrast, at $\theta=9.43^\circ$ the bias gap is nearly closed and $T\approx1$ over a broad $\Delta U$ range, this behavior is interrupted only by a narrow dip in the transmission, occurring when the barrier height aligns with a van Hove singularity. At this point, near-ideal Klein tunneling is observed, with a single resonance similar to those found in large-angle TBLG studies \cite{li2010observation}. 
In the higher energy panels $E/E_V=0.4$ eV (see Figs. \ref{T3} (d–f)), the same patterns remain. Still, the oscillation amplitudes decrease, and their amplitudes shift to larger $\Delta U$, showing that increasing the twist angle and increasing the incident energy reduces the intensity of the resonant pattern. Compared to the unbiased case where transmission is strictly zero for $\Delta U<E_g/2$ and then oscillates once that critical value is exceeded, the applied bias prevents complete suppression at low $\Delta U$ and shifts all resonances toward $\Delta U\approx0$ \cite{he2013chiral}, giving a direct electric field control over the resonance energies, amplitudes, and distribution in TBLG.
\section{Conclusion}\label{S3}
 In this paper, we have studied the chiral tunneling in twisted bilayer graphene (TBLG) under a perpendicular interlayer bias, using a continuum model with dual-gate control over doping and layer potential imbalance. Our calculations, performed across a range of twist angles, reveal that a transverse bias significantly modifies the Klein tunneling behavior in TBLG. In the low-energy regime, the bias induced band gap suppresses perfect transmission at normal incidence, leading to a reflection coefficient approaching unity. However, just above the gap barrier, this suppression is due to twist-dependent miniband states that allow finite tunneling, restoring finite transmission even for near normal incidence.  Furthermore, we find that the interlayer bias breaks the symmetry between the two graphene layers, resulting in pronounced angular and valley asymmetries in the tunneling probabilities. Unlike the unbiased case where transmission remains symmetric for electrons incident at angles $\pm \theta$, the biased TBLG exhibits direction selective tunneling, where electrons approaching from opposite sides of the barrier transmit with different probabilities.
Additionally, valley selective tunneling occurs, breaking the equivalence between the $K$ and $K'$ valleys. Finally, we demonstrate that the interlayer bias modulates Fabry-Pérot-like resonances in the transmission spectrum. The energies and widths of the interference fringes change when a bias is applied, presenting an extra key to control quantum interference effects in electron transport.  
\bibliographystyle{apsrev4-2}      
\bibliography{mybib}
\appendix\label{App.A}
\begin{widetext}
\section{Analytical derivation of the transfer matrix and transmission Probability}
The boundary conditions for the wavefunction and its derivative at the interfaces of the biased TBLG structure result in the following system
\begin{align}
  & r + d_1 - a_2 - b_2 - c_2 - d_2 = -1, \label{eq:1} \\
  & -i k_{x1} r + k_{x2} d_1 - i q_{x1} a_2 + i q_{x1} b_2 + q_{x2} c_2 - q_{x2} d_2 = -i k_{x1}, \label{eq:2} \\
  & s_1 \rho_1^{-} r + s_1 \rho_2^{-} d_1 - s_2 \zeta_1^{+} a_2 - s_2 \zeta_1^{-} b_2 - s_2 \zeta_2^{+} c_2 - s_2 \zeta_2^{-} d_2 = -s_1 \rho_1^{+}, \label{eq:3} \\
  & -i k_{x1} s_1 \rho_1^{-} r + k_{x2} s_1 \rho_2^{-} d_1 - i q_{x1} s_2 \zeta_1^{+} a_2 + i q_{x1} s_2 \zeta_1^{-} b_2 + q_{x2} s_2 \zeta_2^{+} c_2 - q_{x2} s_2 \zeta_2^{-} d_2 = -i k_{x1} s_1 \rho_1^{+}, \label{eq:4} \\
  & e^{i q_{x1} D} a_2 + e^{-i q_{x1} D} b_2 + e^{-q_{x2} D} c_2 + e^{q_{x2} D} d_2 - e^{i k_{x1} D} t - e^{-k_{x2} D} c_3 = 0, \label{eq:5} \\
  & i q_{x1} e^{i q_{x1} D} a_2 - i q_{x1} e^{-i q_{x1} D} b_2 - q_{x2} e^{-q_{x2} D} c_2 + q_{x2} e^{q_{x2} D} d_2 - i k_{x1} e^{i k_{x1} D} t + k_{x2} e^{-k_{x2} D} c_3 = 0, \label{eq:6} \\
  & s_2 \zeta_1^{+} e^{i q_{x1} D} a_2 + s_2 \zeta_1^{-} e^{-i q_{x1} D} b_2 + s_2 \zeta_2^{+} e^{-q_{x2} D} c_2 + s_2 \zeta_2^{-} e^{q_{x2} D} d_2 - s_1 q_1^{+} e^{i k_{x1} D} t - s_1 q_2^{+} e^{-k_{x2} D} c_3 = 0, \label{eq:7} \\
  & i q_{x1} s_2 \zeta_1^{+} e^{i q_{x1} D} a_2 - i q_{x1} s_2 \zeta_1^{-} e^{-i q_{x1} D} b_2 - q_{x2} s_2 \zeta_2^{+} e^{-q_{x2} D} c_2 + q_{x2} s_2 \zeta_2^{-} e^{q_{x2} D} d_2 - i k_{x1} s_1 q_1^{+} e^{i k_{x1} D} t + k_{x2} s_1 q_2^{+} e^{-k_{x2} D} c_3 = 0. \label{eq:8}
\end{align}
From Eqs.~\eqref{eq:1}--\eqref{eq:4}, the left interface matrix $\mathbf{I}_L $ couples $r, d_1 $ to $ a_2, b_2, c_2, d_2 $
\begin{equation}
\mathbf{I}_L = \begin{pmatrix}
1 & 1 & -1 & -1 & -1 & -1 & 0 & 0 \\
-i k_{x1} & k_{x2} & -i q_{x1} & i q_{x1} & q_{x2} & -q_{x2} & 0 & 0 \\
s_1 \rho_1^{-} & s_1 \rho_2^{-} & -s_2 \zeta_1^{+} & -s_2 \zeta_1^{-} & -s_2 \zeta_2^{+} & -s_2 \zeta_2^{-} & 0 & 0 \\
-i k_{x1} s_1 \rho_1^{-} & k_{x2} s_1 \rho_2^{-} & -i q_{x1} s_2 \zeta_1^{+} & i q_{x1} s_2 \zeta_1^{-} & q_{x2} s_2 \zeta_2^{+} & -q_{x2} s_2 \zeta_2^{-} & 0 & 0 \\
\end{pmatrix}.
\end{equation}
The propagation matrix $\mathbf{P}$  accounts for phase evolution and decay in the barrier region
\begin{equation}
\mathbf{P} = \begin{pmatrix}
e^{i q_{x1} D} & 0 & 0 & 0 & 0 & 0 & 0 & 0 \\
0 & e^{-i q_{x1} D} & 0 & 0 & 0 & 0 & 0 & 0 \\
0 & 0 & e^{-q_{x2} D} & 0 & 0 & 0 & 0 & 0 \\
0 & 0 & 0 & e^{q_{x2} D} & 0 & 0 & 0 & 0 \\
0 & 0 & 0 & 0 & e^{i k_{x1} D} & 0 & 0 & 0 \\
0 & 0 & 0 & 0 & 0 & e^{-k_{x2} D} & 0 & 0 \\
0 & 0 & 0 & 0 & 0 & 0 & 1 & 0 \\
0 & 0 & 0 & 0 & 0 & 0 & 0 & 1 \\
\end{pmatrix}.
\end{equation}
From Eqs.~\eqref{eq:5}--\eqref{eq:8}, the right interface matrix $ \mathbf{I}_R $ couples $ a_2, b_2, c_2, d_2$ to $ t, c_3 $
\begin{equation}
\mathbf{I}_R = \begin{pmatrix}
e^{i q_{x1} D} & e^{-i q_{x1} D} & e^{-q_{x2} D} & e^{q_{x2} D} & -e^{i k_{x1} D} & -e^{-k_{x2} D} & 0 & 0 \\
i q_{x1} e^{i q_{x1} D} & -i q_{x1} e^{-i q_{x1} D} & -q_{x2} e^{-q_{x2} D} & q_{x2} e^{q_{x2} D} & -i k_{x1} e^{i k_{x1} D} & k_{x2} e^{-k_{x2} D} & 0 & 0 \\
s_2 \zeta_1^{+} e^{i q_{x1} D} & s_2 \zeta_1^{-} e^{-i q_{x1} D} & s_2 \zeta_2^{+} e^{-q_{x2} D} & s_2 \zeta_2^{-} e^{q_{x2} D} & -s_1 q_1^{+} e^{i k_{x1} D} & -s_1 q_2^{+} e^{-k_{x2} D} & 0 & 0 \\
i q_{x1} s_2 \zeta_1^{+} e^{i q_{x1} D} & -i q_{x1} s_2 \zeta_1^{-} e^{-i q_{x1} D} & -q_{x2} s_2 \zeta_2^{+} e^{-q_{x2} D} & q_{x2} s_2 \zeta_2^{-} e^{q_{x2} D} & -i k_{x1} s_1 q_1^{+} e^{i k_{x1} D} & k_{x2} s_1 q_2^{+} e^{-k_{x2} D} & 0 & 0 \\
\end{pmatrix}.
\end{equation}
The total transfer matrix is
\begin{equation}
\mathbf{M} = \mathbf{I}_R \cdot \mathbf{P} \cdot \mathbf{I}_L, \label{eq:composite}
\end{equation}
and the system $ \mathbf{M} \mathbf{x} = \mathbf{b} $ with 
\begin{equation}
\mathbf{x} = [r, d_1, a_2, b_2, c_2, d_2, t, c_3]^T,\quad  \mathbf{b} = [-1, -i k_{x1}, -s_1 \rho_1^{+}, -i k_{x1} s_1 \rho_1^{+}, 0, 0, 0, 0]^T .
\end{equation}
The transmission $t$ and reflection $r$ coefficients are derived from Eq.~\eqref{eq:composite} via block‐matrix inversion
\begin{equation}
t = -\frac{
  \det\begin{pmatrix}
  \mathbf{M}_{1:4,1:2} & \mathbf{M}_{1:4,5:6} \\
  \mathbf{M}_{5:8,1:2} & \mathbf{M}_{5:8,5:6}
  \end{pmatrix}
}{
  \det\begin{pmatrix}
  \mathbf{M}_{1:4,3:4} & \mathbf{M}_{1:4,5:6} \\
  \mathbf{M}_{5:8,3:4} & \mathbf{M}_{5:8,5:6}
  \end{pmatrix}
},\quad r = -\frac{
  \det\begin{pmatrix}
  \mathbf{M}_{1:4,5:8} & \mathbf{M}_{1:4,2:8} \\
  \mathbf{M}_{5:8,5:8} & \mathbf{M}_{5:8,2:8}
  \end{pmatrix}
}{
  \det\begin{pmatrix}
  \mathbf{M}_{1:4,1:4} & \mathbf{M}_{1:4,5:8} \\
  \mathbf{M}_{5:8,1:4} & \mathbf{M}_{5:8,5:8}.
  \end{pmatrix}
}
\end{equation}
Here, $\mathbf{M}_{i:j,k:\ell}$ denotes the submatrix formed by rows $i$ through $j$ and columns $k$ through $\ell$. 
After straightforward algebra, the transmission probability becomes
\begin{equation}
T = \frac{4\,k_{x1}^2\,q_{x1}^2}
{(k_{x1}^2 + q_{x1}^2)^2 \sinh^2\bigl(q_{x2}D\bigr)
 \;+\;4\,k_{x1}^2\,q_{x1}^2 \cosh^2\bigl(q_{x2}D\bigr)}.
\end{equation}
Similarly, the reflection probability reads
\begin{equation}
R = \frac{(k_{x1}^2 - q_{x1}^2)^2 \sinh^2\bigl(q_{x2}D\bigr)}
{(k_{x1}^2 + q_{x1}^2)^2 \sinh^2\bigl(q_{x2}D\bigr)
 \;+\;4\,k_{x1}^2\,q_{x1}^2 \cosh^2\bigl(q_{x2}D\bigr)}.
\label{eq:R}
\end{equation}
\end{widetext}
\end{document}